\documentclass[10pt,conference]{IEEEtran}
\IEEEoverridecommandlockouts

\usepackage{cite}
\usepackage{amsmath,amssymb,amsfonts}
\usepackage{algorithmic}
\usepackage{graphicx}
\usepackage{textcomp}
\usepackage{xcolor}
\def\BibTeX{{\rm B\kern-.05em{\sc i\kern-.025em b}\kern-.08em
    T\kern-.1667em\lower.7ex\hbox{E}\kern-.125emX}}
    
\usepackage[utf8]{inputenc}
\usepackage{booktabs}
\usepackage{enumitem}
\usepackage{graphicx}
\usepackage{hyperref} 
\usepackage{todonotes}
\usepackage{dirtree,array}
\usepackage{lipsum}
\usepackage{tabularx}
\usepackage{listings}

\lstdefinestyle{mystyle}{
  frame = single, 
  basicstyle=\ttfamily\tiny,
  breakatwhitespace=false,         
  breaklines=true,                 
  captionpos=b,                    
  keepspaces=true,                 
  numbers=left,                    
  numbersep=5pt,                  
  showspaces=false,                
  showstringspaces=false,
  showtabs=false,                  
  tabsize=2,
  title={}
}
\begin{document}

\title{Towards identifying and minimizing customer-facing documentation debt\\
\thanks{Ericsson AB, KKS foundation}
}

\author{\IEEEauthorblockN{Lakmal Silva}
\IEEEauthorblockA{\textit{Department of Software Engineering} \\
\textit{Blekinge Institute of Technology and Ericsson AB}\\
Karlskrona, Sweden \\
lakmal.silva@bth.se}
\and
\IEEEauthorblockN{Michael Unterkalmsteiner}
\IEEEauthorblockA{\textit{Department of Software Engineering} \\
\textit{Blekinge Institute of Technology}\\
Karlskrona, Sweden \\
michael.unterkalmsteiner@bth.se}
\and
\IEEEauthorblockN{Krzysztof Wnuk}
\IEEEauthorblockA{\textit{Department of Software Engineering} \\
\textit{Blekinge Institute of Technology}\\
Karlskrona, Sweden \\
krzysztof.wnuk@bth.se}
}

\maketitle

\begin{abstract}
\textit{\textbf{Background:}} Software documentation often struggles to catch up with the pace of software evolution. The lack of correct, complete, and up-to-date documentation results in an increasing number of documentation defects which could introduce delays in integrating software systems. In our previous study on a bug analysis tool called MultiDimEr, we provided evidence that documentation-related defects contribute to a significant number of bug reports. 
\textit{\textbf{Aims:}} First, we want to identify documentation defect types contributing to documentation defects and thereby identifying documentation debt. Secondly, we aim to find pragmatic solutions to minimize most common documentation defects to pay off the documentation debt in the long run.
\textit{\textbf{Method:}} We investigated documentation defects related to an industrial software system. First, we looked at the types of different documentation and associated bug reports. We categorized the defects according to an existing documentation defect taxonomy.
\textit{\textbf{Results:}} Based on a sample of 101 defects, we found that a majority of defects are caused by documentation defects falling into the Information Content (What) category (86). Within this category, the documentation defect types Erroneous code examples (23), Missing documentation (35), and Outdated content (19) contributed to most of the documentation defects. We propose to adapt two solutions to mitigate these types of documentation defects. 
\textit{\textbf{Conclusions:}} In practice, documentation debt can easily go undetected since a large share of resources and focus is dedicated to deliver high-quality software. This study provides evidence that documentation debt can contribute to increase in maintenance costs due to the number of documentation defects. We suggest to adapt two main solutions to tackle documentation debt by implementing (i) Dynamic Documentation Generation (DDG) and/or (ii) Automated Documentation Testing (ADT), which are both based on defining a single and robust information source for documentation. 
\end{abstract}

\begin{IEEEkeywords}
Documentation Debt, Technical Debt, Automation
\end{IEEEkeywords}

\section{Introduction}
\par
As software development is a human oriented task~\cite{lethbridge2003software}, the documentation of software becomes a vital interface between the software system and its user and developers. A well documented and an up-to-date documentation provides a better understanding of the system in various phases of the software development and maintenance cycles~\cite{lethbridge2003software}. However, prior research~\cite{venigalla2021understanding, zhi2015cost} and our industry experience show that maintaining high-quality documentation is seldom prioritized. Software documentation is often treated as a second class artifact and is managed as an afterthought within the software development process~\cite{aghajani2020software}. 
\par
Different types of software documentation are produced during software development, such as requirements documents, test documents, developer documents, code comments, and end-user documents~\cite{venigalla2021understanding} to name a few. The end user documents or the customer facing-documents, the focus in this paper, are a crucial set of documents that are produced to be used by users internal or external to the product development organization. Customer-facing documentation is usually the entry point to understand, install, and manage a software system. As opposed to requirements documents, test documents, developer documentation, and code comments, these documents are an integral part of a software system and are version controlled and delivered together with a software system. Therefore, customer facing-documents can contribute to defects and technical debt accumulation, which deserves attention similar to the technical debt management of software artifacts.
\par
Software defects consume a significant amount of time and money~\cite{dehaghani2013factors} for both the development organizations as well the end users. In an effort to identify Technical Debt (TD), we implemented a bug analysis tool called MultiDimEr, a Multi-Dimensional bug analyzEr~\cite{silva2022multidimer} that analyzes and categorizes bug reports into different dimensions such as architectural components, source code files, and customer-facing documents. The analysis tool revealed that most of the reported defects resulted in updates to customer-facing software documents such as configuration guides, deployment guides and user guides. This revelation prompted us to investigate documentation debt, which has never been a focused area in the software development organization that we worked with at Ericsson. We identified two main causes for documentation updates due to defects. The first one is related to insufficient and inadequate content and obsolete, ambiguous information, as also pointed out by a survey conducted by Aghajani et al.~\cite{aghajani2020software}. The second cause are source code defect fixes such as installation, upgrade/migration scripts that require documentation updates. 
\par
The goal of our research is to identify causes for documentation debt in customer-facing documents and find solutions to minimize such debt. Certain types of customer-facing documents such as Deployment Guides, Installation Guides, and API References consist of a combination of natural language text and command syntax, whereas documents such as User Manuals and Getting Started Guides vastly consist of descriptive natural language text. Hence, the documentation defect types and thereby the solutions to tackle documentation debt can vary.
\par
 We narrowed down our solutions to cover the defect types Erroneous code examples, Missing documentation and Outdated content, as our analysis showed that these types caused most of the documentation bug reports. The \textbf{main contributions} of this research are:
\begin{itemize}
  \item A method for identifying documentation debt from bug reports with the help of a documentation defect taxonomy.
  \item Further empirical validation of the documentation defect taxonomy in the context of documentation debt.
  \item A description of solutions to the most common documentation defects contributing to documentation debt.
\end{itemize}
\par
The rest of this paper is structured as follows. Section~\ref{relatedwork} provides an overview of prior research on documentation debt and proposed solutions. We describe our research design in Section~\ref{ResDes}, including the Research Questions (RQs). Section~\ref{results} reports the results from our investigation, followed by adapting two solution proposals in Section~\ref{solutions} to mitigate the identified common documentation defects. We discuss the results of our investigation in Section~\ref{discussion}. We conclude our paper and provide directions for future work in Section~\ref{conclusion}.

\section{Related work}\label{relatedwork}
\par
Technical Debt (TD) in Software Engineering is a widely researched area that has even expanded to more fine grained TD types~\cite{alves2014towards} such as architectural debt~\cite{besker2018managing, verdecchia2018architectural}, code debt~\cite{amanatidis2018developer, fontana2015towards}, test debt~\cite{aragao2022testdcat, samarthyam2017understanding}, and documentation debt~\cite{kruchten2012technical}. The term TD was coined by Cunningham~\cite{cunningham1992wycash} in 1992 referring to sub-optimal decisions/implementations taken to meet short term benefits that contribute to costs in the long run due to limitations in evolving and maintaining the system. 
\par
Documentation debt, which is the focus of this study, refers to missing, inadequate or incomplete documentation~\cite{tom2013exploration, seaman2011measuring, alves2014towards}. A characteristic of TD is that it is usually visible in the quality aspects of a product, but mostly invisible in the artifacts of a product, like design, source code and tests~\cite{kruchten2012technical}. The software industry has progressed in identifying certain types of TD such as the source code and test debt by adding instrumentation to analyze source code~\cite{pina2022sonarlizer} through tools such as SonarQube\footnote{\url{https://www.sonarsource.com/products/sonarqube/}} and PMD\footnote{\url{https://pmd.github.io/}} that can expose the hidden TD to developers. However, we have not encountered similar tools for identifying documentation debt in practice. One way to overcome this limitation is to study defect reports associated with documentation artifacts. They can be a signal of documentation debt and analysing their distribution and frequency can provide insights on where the debt occurs. Furthermore, an analysis would allow to make informed decisions on whether it would make sense to attempt to prevent the debt instead of paying the principal in form of fixing defects and the impression of low product quality at the customer. 
\par
Codabux et al.~\cite{codabux2021technical} studied TD in scientific software. They analyzed peer-review comments of packages that were submitted to a repository collecting scientific R packages\footnote{\url{https://www.r-project.org/}}. They manually classified 358 comments originating from 157 packages and created a taxonomy of ten technical debt types. They found that documentation debt was the most prominent, with close to 30\% of all found instances of TD. The predominance of documentation debt was further substantiated by Khan and Uddin~\cite{khan2022automatic} who automated the classification and analyzed 13.500 comments originating from 1297 packages. Looking at the taxonomy proposed by Codabux et al.~\cite{codabux2021technical}, they define documentation debt as deficits in code documentation, as well as build and end-user documentation. In this paper, we focus on customer-facing documentation as we found that this type of documentation contains the most defects in the system we studied~\cite{silva2022multidimer}, further substantiating that documentation debt is the most frequently encountered type of TD. 
\par
Aghajani et al.~\cite{aghajani2019software} focused their investigation on developing a more differentiated categorisation of documentation debt. They mined a large collection of documentation related data sourced from discussions on StackOverflow, issues and pull requests on GitHub, and mailing lists from the Apache Software Foundation. The resulting hierarchical taxonomy, which we also use in this research, contains 162 documentation defect types that are relevant for software developers.  
\par
To address absent or outdated documentation, prior research has proposed the auto generation of documentation through source code summarization methods~\cite{moreno2013automatic, mcburney2014automatic, mcburney2015automatic}, and more recently, to produce on-demand documentation~\cite{robillard2017demand}. However, most of these solutions are targeting developer documentation, which is different to customer-facing documentation in terms of the target audience, the documentation content and how they have been produced. For instance, developer documentation is internal raw documentation whereas customer-facing documentation is external and formatted to be used by the external users of the system~\cite{raglianti2022topology}. Developer documentation is usually maintained by developers whereas customer-facing documentation is written and maintained by technical writers~\cite{raglianti2022topology} that follows different tools and processes compared to loosely managed developer documentation.
\par
Another interesting approach is executable documentation~\cite{tegeler2022executable} where domain-specific notations are turned into fully-fledged modelling/programming languages, or more specifically, domain specific languages (DSLs). There is a relation between documentation with software models as argued by Stevens~\cite{stevens2022models}, where models can be used to document software while in some cases, the documentation can be used to generate models. However, these approaches are still in their initial stages and require further research to be used in practice~\cite{tegeler2022executable}.
\par
Another related approach to minimize documentation defects is automatic documentation testing/verification. The DASE (Document-Assisted Symbolic Execution) approach~\cite{wong2015dase} suggests the use of program documentation to extract input constrains for testing. Another tool called DScribe uses a mechanism to combine unit tests and documentation through templates that are used to generate documentation and unit tests~\cite{nassif2021generating}. A software verification and a functional testing method for machine interpreted documentation was introduced by Friedman-Hill et al.~\cite{friedman2001software}, by incorporating documentation testing to a test framework.
\par
We were unable to find prior studies targeted at systematically identifying customer-facing documentation debt. Hence, we aim to fill this gap by proposing and testing a customer-facing documentation debt identification method using bug reports and a  documentation taxonomy in an empirical context. We also aim to contribute with adapting pragmatic solutions based on the identified debt in the customer-facing documentation context.

\section{Research Design}\label{ResDes} 
\par
The aim of this research is to identify the causes of documentation debt and to investigate possible solutions to minimize documentation defects in the future.
We embed this aim in the context of a particular product (referred to as System A), developed at Ericsson.
System A is currently under active development and has already been released in multiple versions to the market. The current version of the product is built on a microservices architecture and is deployed on the cloud native platform  Kubernetes~\footnote{\url{https://kubernetes.io/}}. The life-cycle of the application is managed by Helm~\footnote{\url{https://helm.sh/}}, which is a package management system and a life-cycle management system for Kubernetes.

Our analysis provides a chain of evidence related to documentation defects, which can be used to motivate the required investments in documentation improvement solutions. 

\subsection{Research Questions}

We define the following research questions (RQs) to guide our investigation.

\begin{enumerate}[label=RQ\arabic*]
    \item \label{itm:RQdoctypes}What are the types of customer-facing documents that contain most of the defects?
    \par
    We are interested in understanding whether certain types of documents contain more defects than others. This would allow us to narrow down the design of a solution, which likely needs to be adapted to the particular document type.
    \item \label{itm:RQdocerrors}What type of customer-facing documentation defects can be observed in bug reports?
    \par
    The goal of this RQ is to understand if defects are due to accumulated documentation debt or due to random and ad-hoc documentation defects. To identify and quantify documentation debt, we use bug reports and a documentation defect taxonomy introduced by Aghajani et al.~\cite{aghajani2019software} to group related defects. Even though Aghajani et al. validated the documentation defect taxonomy with practitioners~\cite{aghajani2020software}, the results can be subjective due to personal opinions since the study was conducted through surveys. We complement the taxonomy by validating it with further empirical data.
    \par
    To the best of our knowledge, Aghajani et al.'s taxonomy has not been used for documentation debt analysis before. Hence, a related sub question is:
    \begin{enumerate}[label=RQ2.\arabic*]
        \item \label{itm:RQtaxSupport} To what extent does the taxonomy support the classification of customer-facing documentation defects in industry?
    \par We believe that the taxonomy is useful for documentation defect classification and quantification, which is a key element in identifying documentation debt. Since the taxonomy is being used for the first time to identify documentation debt, we reflect on how well the taxonomy fits for this purpose.
    \end{enumerate}
    \item \label{itm:RQdocCost} What is the cost of the customer-facing documentation defects?
    \par
    It is necessary to estimate the cost of documentation defects to motivate the benefits of paying off documentation debt by implementing the proposed solutions.
    \item \label{itm:RQsolution}How can we minimize the customer-facing documentation defects through automation?
    \par
    Based on the quantification of documentation defects identified as part of \ref{itm:RQdocerrors}, we are identifying and describing solutions to mitigate the most common defect causes.

   % \item What is the effort required to minimize the identified doc TD? [NOTE: Not sure if this RQ is needed ]
\end{enumerate}

\subsection{Data collection}\label{dataCollection}
We utilized MultiDimEr to collect and classify bug reports submitted between March 2019 and September 2022, belonging to System A that are stored in a central bug management system. The first bug report of the cloud native version of System A was reported in March 2019, hence we used it as the starting point for data collection. This data set contains a total of 1663 resolved bug reports where 438 bug reports resulted in documentation updates according to MultiDimEr's classification.  Out of the 438 bug reports, 120 bug reports resulted in documentation updates due to source code changes. The remaining 318 bug reports targeted issues purely originating in documentation defects. Hence, we focus our analysis on this set of defects. 

\subsection{Data Analysis}
\par
 We used a sample study strategy, as suggested by Stol et al.~\cite{stol2018abc}, to achieve generalizability of documentation defect types in the context of System A from a sample of bug reports. The overall bug reports analysis consists of three steps as outlined below.
 
\par
The first step is to understand the distribution of bug reports over different documents for answering~\ref{itm:RQdoctypes}. The bug reports distribution was obtained via the classification results from MultiDimEr. We observed documentation updates as part of software defects. However, for the scope of this study, we only considered pure documentation defects.

\par
In the second step, we classified documentation defects to one or more documentation defect categories from  Aghajani et al.'s taxonomy. A sample of 101 from a total of 318 bug reports related to 67 customer-facing documents was used. We selected a representative bug report sample by including at least one bug report from each document.  From the documents that contained the majority of bug reports, we included at least half of the bug reports into the sample. The outcome from this analysis helps us to answer~\ref{itm:RQdocerrors}. The protocol that we used for the classification is described below: 

\begin{enumerate}
    \item Select a bug report.
    \item Read the observation and the answer sections of the bug report.
    \item Based on the information from the observation and the answer sections, classify the defect to one or more sub categories within the main information content types ``What'' and/or ``How'' of the taxonomy. The defects within information content type ``What'' refer to  ``issues arising from \emph{what} is written in the documentation''~\cite{aghajani2019software} and the defects within information content type ``How'' refer to ``\emph{how} the content is written and organized''~\cite{aghajani2019software}.
\end{enumerate}

\par
Although the taxonomy introduced by Aghajani et al. contains four top level categories (``What'', ``How'', ``Process Related'' and ``Tools Related''), we decided to use only the ``What'' and ``How'' categories. Aghajani et al. \cite{aghajani2019software} reported that ``What'' (485 defects), ``How'' (255) type defects are more frequent compared to  ``Process Related'' (81) and ``Tools Related'' (134) defects, so we conjectured that the majority of the bug reports can be covered by these two categories. Too many categories makes the classification difficult, as defects may get distributed into overlapping defect categories, making the defect frequency distribution less useful to derive conclusions. 

\par
The third step involved the analysis of the most frequent documentation defect categories, motivated by the rationale that the development and application of a solution should address the most frequently encountered defects. The description of the identified solutions answers~\ref{itm:RQsolution}.

\subsection{Piloting the bug report classification}
We conducted a pilot classification to test the accuracy and the efficiency of the classification protocol. We selected 10 bug reports from a document called ``Configuration Management'' belonging to System A, which contained most of the documentation bug reports. The lead author classified 10 bug reports while the second and the third authors classified five each. Eight out of the 10 bug reports were classified to the same documentation defect category by two persons. We extended the pilot classification with five more bug reports on the ``Installation Guide'' of System A. This extension of the pilot was to get an affirmation that the classification protocol can be applied independently of the documentation types. From the extended pilot we observed that four out of the five bug reports were classified to the same documentation categories. The agreement between the three authors provides us confidence on the repeatability of the classification results on any documentation type. On average we spent around three minutes per bug report analysis, which is reasonable enough to scale the classification to a larger sample.

\subsection{Defect prioritization and Cost estimates} \label{sec:defectCost}
\par
Unlike source code bug reports, the documentation bug reports are in most circumstances registered with a lower severity, as they usually do not affect the core business functionality. However, certain defects can be of a high priority, for example the defects detected by the external users, defects on documents such as application programming interfaces (APIs), installation and configuration guides.

We calculated the time between the bug report registration and assignment to a developer. This can be used as a defect severity and a prioritization indicator. The time period between registration and assignment indicates relative priority. 
\par
Like any other defect, documentation defects also incur significant costs on different levels. Hence we need a mechanism to estimate such costs. To start with, the users of documents spend time troubleshooting the issues when things do not work as they are documented, and report them by creating bug reports. Once a bug report is received by the development organization, costs incur as part of management activities such as bug assignment, documentation fixes, documentation verification and sending out correction packages for document collections. 
Since System A did not have a cost estimation framework within defect management, we use three defect report variables to approximate the documentation defect cost:
\begin{itemize}
    \item The proportion of internally to externally detected defects. The rationale is that defects detected by customers are more costly to fix and have also detrimental side effects, like loss of confidence in the product.
    \item The severity of defects assigned by the bug reporter.
    \item The time between a bug report assignment until the bug fix was accepted, approximating the cost of resolution. A longer time period between bug registration and solution acceptance is an indicator that the defect may be complex to be handled and may incur higher costs.
\end{itemize}

We updated the MultiDimEr tool to collect this extra information.

\subsection{Threats to validity}
There can be a variety of threats to validity when conducting empirical research. However, we have taken steps to minimize those treats to the best of our ability, which are outline below.

\par
Manual classification by humans can be subjected to bias. We mitigated this threat by piloting the classification of bug reports into documentation defect categories among the three authors to understand how reliably the classification can be conducted independently and how much of agreement exists between independent classifications. To minimise subjectivity in the classification, we also annotated the text from the bug reports that led to the chosen classification, allowing us to identify the root causes for potential disagreement and align our common understanding of the defect categories. 

\par
When dealing with empirical studies, there may be threats to external validity as different companies have different ways of working, different development processes and more importantly, how the customer-facing documentation is named, structured and managed. To minimize external validity, regarding documentation naming, we mapped the Ericsson documentation into more generic and already used naming conventions~\cite{aghajani2020software} from prior research. 

\par
We have studied documentation defects in a specific industry context. Hence, we took precaution to describe first the concepts behind the solutions so they can be adapted and implemented in different contexts. In addition, we provide technology specific implementation details according to our chosen industrial system, that can be beneficial for practitioners that use similar technologies. 

\par
The industrial system that we investigated is a cloud native system that is deployed in Kubernetes environments. The documentation that contained most of the defects is thematically connected to the platform. However, our findings and solutions are not platform dependent.

\section{Results}\label{results}
We answer~\ref{itm:RQdoctypes}, i.e. the bug report distribution among different document types, with the results from the analysis performed by the MultiDimEr. The exact naming of the documents is irrelevant outside the Ericsson context and use therefore the documentation categories introduced by Aghajani et al.~\cite{aghajani2020software}. A total of 67 customer-facing documents from System A were grouped into six categories: API References, Getting Started Guides, Installation Guides, Deployment Guides, Release Notes/Change Logs and User Manuals. Table~\ref{table:docdefects} illustrates the distribution of the bug reports. From the results we see that just over half of the issues were reported on the Deployment Guides (129) and the Installation Guides (53). It is worth highlighting that the dynamically generated Release Notes only contained eight issues. Over the years, the management decided to invest in dynamically generating the Release Notes to shorten delivery preparation time and support continuous deliveries. This document contains information such as added new features, corrected bug report information, and microservices version information. We could conjecture that the lower number of bugs is due to the dynamic documentation generation from a robust information source.
\par

\begin{table}[htbp]
  \caption{Bug reports distribution among various document types}
  \label{table:docdefectsdistro}
  \begin{center}
  \begin{tabular}{p{0.35\linewidth} p{0.4\linewidth}}
    \toprule
     Document Type & No. of bug reports \\
      \midrule
      Deployment Guides & 129 \\
      Installation Guides & 53 \\
      API References & 51 \\
      User Manuals & 50 \\
      Getting Started Guides &  27\\
      Release Note/Change Log & 8 \\
      \midrule
      Total documentation bugs &  \textbf{318} \\      
      \bottomrule
  \end{tabular}
  \end{center}
\end{table}

\par
Next, we report results from our classification of the bug reports to the documentation defect taxonomy, answering \ref{itm:RQdocerrors}. From the results in Table~\ref{tab:classficationResults}, we can observe that 85\% (86 out of 101) of the defects fall into the Information Content (What) category. These 86 issues are distributed among the second level of issues: Completeness (37), Correctness (30), and Up-to-dateness (19). On the third level of defect categories, the defects were mostly distributed between on Erroneous code examples (23), Missing configuration instructions (14), Missing/Poor documentation (15), and Outdated content (14) in relation to system evolution. 
\par
A commonality of these third level categories is that they are related to step by step instructions and/or command syntax that were either missing, incorrect or outdated. Following are some examples of defects that we found from the investigated bug reports:

\begin{enumerate}[label=E.g.\arabic*]
  \item ``We need add a "--reuse-values" flag for the command to work''
  \item ``service names are incorrect''
  \item \label{restdesc} ``The configuration to enable the external IP for REST is not described''
  \item \label{stepbystep} ``The Configuration Management and Deployment Guide lacks detailed step-by-step instructions on how to both properly configure service-x''
\end{enumerate}

The most obvious implication from the above defects is the management overhead (defect identification, assignment, and acceptance of the solution) of the documentation bug reports. However there are other implications that are hidden, such as introducing delays to the projects, the cost of troubleshooting and in some cases (in~\ref{restdesc} and \ref{stepbystep} above) the need to call for emergency support which is a very costly activity.

In relation to \ref{itm:RQtaxSupport}, the taxonomy indeed helped us to categorize the documentation defects to the first and the second level categories easily. However, the third level categories in the taxonomy by Aghajani et al.~\cite{aghajani2020software} are highly influenced by source code related or developer documentation. This led to some uncertainty in categorization. For example, we observed many bug reports due to incorrect/outdated commands. However, there is no adequate category in the taxonomy for such defects. The closest was the Erroneous code examples category and the Outdated example category.

Only around 15\% (15 out of 101) of issues were related to Information Content (How) category. Therefore, we only focused on solutions that address issues related to the Information Content (What) category in this study.

\begin{table*}[ht]
\caption{Documentation defects (note that one defect report can be classified into several defect categories)}
\label{tab:classficationResults}
\centering
\begin{tabular}{ccc}
\toprule
Defect categories & Frequency & Solution type \\
\midrule
\begin{minipage}{10cm}\dirtree{%
    .1 \textbf{Information Content (What)}.
    .2 Correctness.
    .3 Erroneous code examples.
    .3 Faulty tutorial.
    .3 Inappropriate installation instructions.
    .2 Completeness.
    .3 Missing configuration instructions.
    .3 Missing unrecommended usage.
    .3 Installation, deployment, \& release.
    .3 Missing code behavior clarifications.
    .3 Other Missing/Poor documentation.
    .2 Up-to-dateness.
    .3 Missing documentation for new feature/component.
    .3 Outdated example.
    .3 Other Up-to-dateness issues.
    .1 \textbf{Information Content (How)}.
    .2 Maintainability.
    .2 Readability.
    .2 Usability.
    .2 usefulness.
    }\end{minipage}    
&  
\DTsetlength{0pt}{0pt}{0pt}{0pt}{0pt}
\begin{minipage}{2cm}\dirtree{%
    .1 86.
    .2 30.
    .3 23.
    .3 4.
    .3 3.
    .2 37.
    .3 14.
    .3 4.
    .3 2.
    .3 2.
    .3 15.
    .2 19.
    .3 7.
    .3 7.
    .3 5.
    .1 15.
    .2 1.
    .2 4.
    .2 7.
    .2 3.
    }\end{minipage}
&
\DTsetlength{0pt}{0pt}{0pt}{0pt}{0pt}
\begin{minipage}{2cm}\dirtree{%
    .1 -.
    .2 -.
    .3 automation.
    .3 automation.
    .3 automation.
    .2 -.
    .3 automation.
    .3 -.
    .3 automation.
    .3 -.
    .3 -.
    .2 -.
    .3 automation.
    .3 automation.
    .3 -.
    .1 -.
    .2 -.
    .2 -.
    .2 -.
    .2 -.
    }\end{minipage}\\
\bottomrule
\end{tabular}
\end{table*}

We report the results related to the cost of documentation defects. As we pointed out in Section~\ref{sec:defectCost}, System A did not have defect cost estimation framework. Hence, we use three quantitative dimensions, derived from the bug report data, to approximate the documentation defect cost: (a) the proportion of internally to externally detected defects, (b) defect severity and (c) time period between bug assignment and bug fix acceptance. Table~\ref{table:docdefects} summarizes the results of (a) and (b).

\begin{table}[htbp]
  \caption{Document bug report cost approximation}
  \label{table:docdefects}
  \begin{center}
  \begin{tabular}{p{0.35\linewidth} p{0.4\linewidth}}
    \toprule
     Bug report dimension & No. of bug reports \\
      \midrule
      Internal &  192 \\
      External  &  126 \\
      \midrule
      Severity C & 299 \\
      Severity B & 19 \\
      Severity A & 0 \\
      \midrule
      Total documentation bugs &  \textbf{318} \\      
      \bottomrule
  \end{tabular}
  \end{center}
\end{table}
The result shows that 40\% of the defects are externally reported  (126 out of 318). However, none of the documentation defects was assigned a high severity (A). This is explainable by considering that only defects that directly affect business operations are assigned a high severity.

The mean time for assigning an internal documentation defect to a team is around 5 days, whereas assigning an external documentation defect takes around 7 days. When an external bug is registered at Ericsson, it needs to be routed between at least two organizations. This routing creates additional delays in reaching the development organization. 

The mean time for resolving an internal documentation defect is around 10 days whereas resolving an external documentation defect is around seven days. This indicates that we may be fixing external bugs at a higher priority once they are being assigned. However, without knowing the effort that went into addressing a defect, this time-based measure is only a weak indicator of defect cost.

The mean time of solving the documentation defects that could have been detected with the proposed two automation solutions (discussed next in Section~\ref{solutions}) is around 11 days. As we have reported in Table~\ref{tab:classficationResults}, this covers 59\% of the documentation defects. Even though we cannot exactly calculate the cost of the defects or the amount of savings, we can get an indication of the relative cost saving that automated solutions to avoid documentation debt would provide: 59\% of 318 are 188 bug reports, which is 11\% of the bug reports (1663) fixed between March 2019 and September 2022 in System A.

\section{Solutions}\label{solutions}
\par
From our analysis of documentation defects, we observed that the majority of the reported defects (59\%) were due to incorrect command syntax, commands related to the product and the platform not being up to date with the software versions, and incorrect execution steps (see Table~\ref{tab:classficationResults}). Minimizing such defects would contribute to the reduction of maintenance costs. Therefore, we limited our solution scope to address the command-related and execution steps related faults. We consider the following key design criteria when identifying the solutions: 

\begin{enumerate}
    \item \textit{Robust information source of the systems behavior.} \label{CR:robustsource}
    One of the root causes for the documentation defects we observed is the lack of a robust information source for documentation. Currently, the input for documentation comes from software developers, who execute commands to test them and forward these to the documentation team. In this manual process, mistakes (wrong assumptions on environment setup, typos) can propagate unnoticed to the documentation as there might be a difference between the commands, executed and verified by the developers, and the documented commands. Additionally, the lack of automated documentation testing makes the information outdated very quickly, since detecting documentation discrepancies is a manual process.
    \item \textit{Automation.} \label{CR:automation}
    The correctness of documentation needs to be verified automatically as the systems evolve and documentation tends to be outdated quickly.
    \item \textit{Developer friendliness.} \label{CR:developer}
    Ericsson has adopted the shift-left concept, which is to move the development, testing and operations of the software system towards production-like systems. Automation in early development phases allows to detect and fix issues as early as possible~\cite{jimenez2019devops}. This entails that the development teams have a greater responsibility to safe guard the quality of the delivered features, including the customer-facing documents. Hence, it is vital to consider the developers when proposing solutions for preventing documentation defects. 
\end{enumerate}

The outlined design criteria have led us to the identification of two solutions that can be adapted: Dynamic Documentation Generation (DDG) and Automated Documentation Testing (ADT). DDG has already been used for summary generation of methods in the source code~\cite{mcburney2015automatic, moreno2013automatic} and API generation~\cite{nybom2018systematic} whereas ADT has been proposed as test-enabled documentation~\cite{nassif2021generating, friedman2001software}. 

In the remainder of this section, we first describe the documentation system, Darwin Information Typing Architecture (DITA), currently used at Ericsson (Section~\ref{sec:dita}). Then, we describe how both DDG (Section~\ref{sec:ddg}) and ADT (Section~\ref{sec:adt}) can be realized with DITA. Based on the requirements at different organizations, a suitable approach can be adapted. In the Ericsson context we chose to adapt DDG, and describe the design in Section~\ref{sec:ddg}.

\subsection{Darwin Information Typing Architecture (DITA)}\label{sec:dita}
\begin{figure}[h]
  \centering
  \includegraphics[width=8cm ,height=3.5cm]{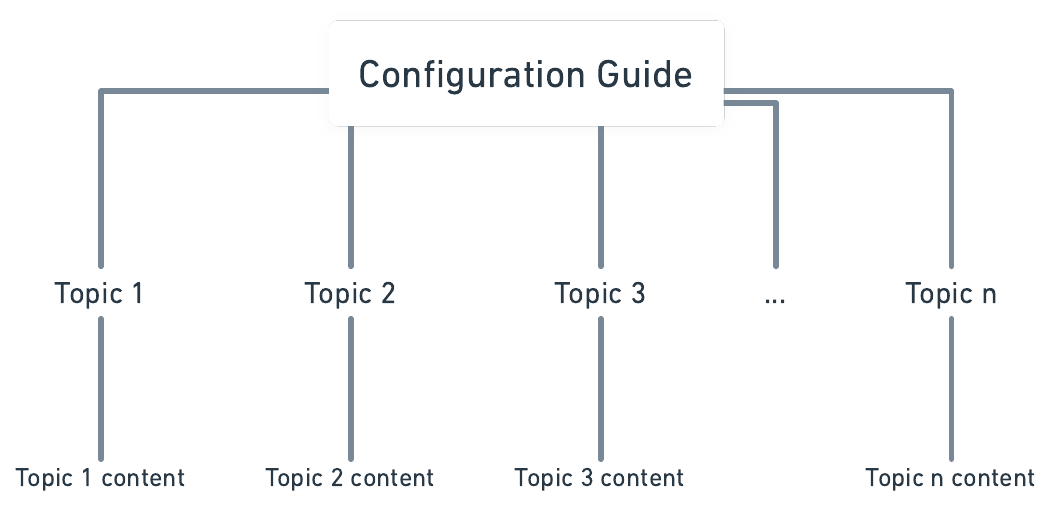}
  \caption{Topic-based organization of DITA documents.}
  \label{Fig:ditastruct}
\end{figure}

The customer-facing documents throughout Ericsson are structured and developed according to the Darwin Information Typing Architecture~\footnote{\url{https://www.oasis-open.org/committees/tc_home.php?wg_abbrev=dita}}. DITA is an open standard that specifies a set of document types for authoring, organizing topic-oriented information. The documents are stored in a format based on the Extensible Markup Language (XML). A key characteristic of DITA based documents is the topic orientation, i.e., a document is composed of smaller sections called topics. A DITA map is used to structure the topics necessary for the document. Figure~\ref{Fig:ditastruct} illustrates how a document called ``Configuration Guide'' consists of multiple topics, while Listing 1 shows an example of the XML based DITA topic. Lines 7-10 render the following command. 

\begin{verbatim}
kubectl get configmap 
<customized_configmap_name> 
-o yaml -n <namespace> 
<customized_configmap_name>-<namespace>.yaml
\end{verbatim}

The parameters within \textlangle\textrangle are to be replaced by actual values based on the site specifications and user requirements.

\begin{figure}[h]
  \centering
  \includegraphics[width=8cm ,height=3cm]{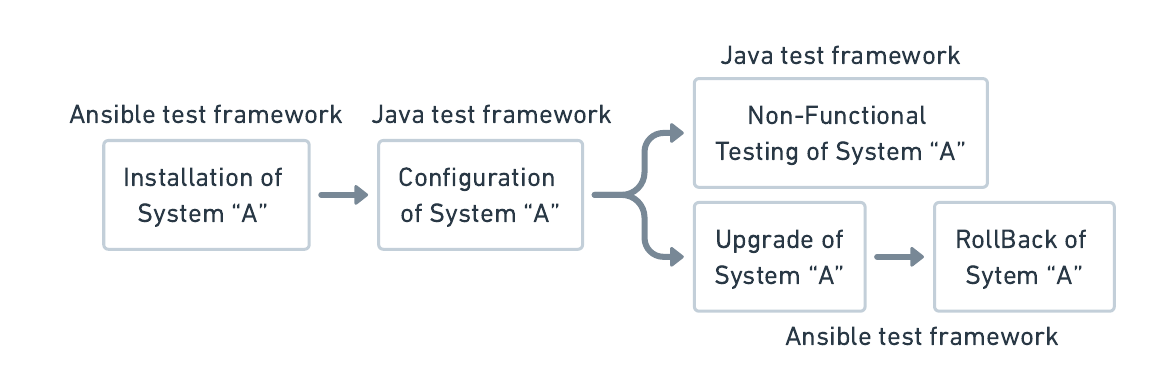}
  \caption{High level overview of testing phases and frameworks.}
  \label{Fig:testprocss}
\end{figure}

The department that develops System A has used customer-facing documents as input when developing the system tests. However, there is no connection between the documentation and the test implementation after the initial development. A high level overview of the testing phases and the test frameworks being used are shown in Figure~\ref{Fig:testprocss}. 
\par
There are two main frameworks to test System A: an Ansible based framework and a Java based framework. The Ansible framework uses the Installation Guides (which also covers the upgrades and rollback procedures). A key characteristic of these documents is that they contain step-wise instructions to execute commands and verify outputs of commands. Ansible is a better fit for Command Line Interface (CLI) based testing. Currently, there are two teams that are  keeping track on the required documentation updates by manually reviewing the source code changes. Additionally, when closer to the releases (three-week cycles), one team is required to manually test the Installation Guides, causing additional one week delay.
\par
On the other hand, the Java framework is influenced by the Deployment Guides, User Manuals and API References, which contain instructions and commands to configure the system in preparation for sending different types traffic. These commands are more complex compared to the CLI commands, hence the use of Java framework. Since its initial implementation using the documentation, there is no monitoring in place to make sure the implemented procedures and the documentation are in sync.  

\subsection{Dynamic Documentation Generation (DDG)}\label{sec:ddg}
\par
Software development organizations nowadays rely on Continuous Integration (CI) and Continuous Delivery (CD) to efficiently deliver software. The CI/CD pipelines use test frameworks that verify different aspects of the software system, such as install/update/upgrade and system configuration. The test code is always evolving and needs to be aligned with the system behaviour. Hence, the test code can be considered a robust information source of the system, fulfilling criterion~\ref{CR:robustsource}. Criterion~\ref{CR:automation} is fulfilled as the test code automatically runs on a daily basis. 
\par
The idea with the DDG approach is to embed meta data in the test code, which can be used to generate the documentation. The test code is closer to the developers so that makes it is easier for them to make the required code changes, which fulfils criterion~\ref{CR:developer}.
\par
The installation/update/upgrade of System A is based on Ansible\footnote{\url{https://www.ansible.com/}} playbooks, so in this study we focused on how Ansible can be used to generate the documentation. Though there exist Ansible modules for generating documentation, they do not fulfill our required criterion of a single robust information source. For instance, the ansible-autodoc\footnote{\url{https://pypi.org/project/ansible-autodoc/}} package uses annotation based document generation but the annotations used for the documentation generation are not used in testing the system. Since the annotations are written as comments, the developers manually still need to keep the comments and the actual commands in sync. 

\par
Listing 2 shows a mock-up of an Ansible playbook code snippet for a new module called dita\_generator that generates the DITA topic snippet shown in Listing 1. When executing this playbook, the commands are executed towards the system under test, while generating the DITA topic snippets required for producing the documentation. 

\renewcommand\lstlistingname{Listing 1: Generated DITA topic from dita\_generator Ansible module }
\renewcommand\lstlistlistingname{mystyle}
\begin{lstlisting}[language=XML]
<section id="section_uzl_2p2_4rb">
    <title>Backup Customized Configmaps and Secrets</title>
    <p>Cluster secrets and configmaps backup are needed if customized configuration was done after installation.</p>
    <ul>
        <li>
            <p>For configmaps, use the following command:</p>
            <userinput>kubectl get configmap <varname>customized_configmap_name</varname> -o yaml -n
                <varname>namespace</varname>
                > <varname>customized_configmap_name</varname>-<varname>namespace</varname>.yaml
            </userinput>   
            </p>
        </li>
        <li>
            <p>For secrets, use the following command:</p>
            <userinput>kubectl get secret <varname>customized_secret_name</varname> -o yaml -n
            <varname>namespace</varname>
            > <varname>customized_secret_name</varname>-<varname>namespace</varname>.yaml
            </userinput>
        </li>
    </ul>
\end{lstlisting}
\renewcommand\lstlistingname{Listing 2: An Ansible playbook snippet illustrating the usage of dita\_generator}
\renewcommand\lstlistlistingname{mystyle}
\begin{lstlisting}[language=TeX]
---
		- name: Test and generate section 2.2 of Config Guide	
			dita_generator:
			  sectionid: section_uzl_2p2_4rb
			  title: Backup Customized Configmaps and Secrets
			  text: Cluster secrets and configmaps backup are needed if customized configuration was done after installation.
			  list:
				text: For configmaps, use the following command:
				userinput: kubectl get configmap {{customized_configmap_name}} -o yaml -n {{namespace}} > {{customized_configmap_name}}-{{namespace}}.yaml
				
				text: For secrets, use the following command:
				userinput: kubectl get secret {{customized_secret_name}} -o yaml -n namespace}} > {{customized_secret_name}}-{{namespace}}.yaml
	
\end{lstlisting}

Implementing a documentation generator for Ansible playbooks which fulfils criterion~\ref{CR:robustsource} requires that the Ansible code written in the playbook is used for both documentation generation as well as generating the commands to be sent towards the System Under Test (SUT).

\subsection{Automated Documentation Testing (ADT)}\label{sec:adt}
\par
Compared to generating documentation from test code, this approach is based on using the existing documentation, and considering it as the robust information source (see criterion~\ref{CR:robustsource}). In this approach, the commands are extracted from the documentation and then used in the test cases that install/update/upgrade the system. In other words, the documented commands are fed into the test cases and are checked when the tests are executed, fulfilling criterion~\ref{CR:automation}. The test results indicate if the system and the corresponding documentation commands are in sync. This approach does not require any developer involvement, hence fulfilling criterion~\ref{CR:developer}.
\par
\begin{figure}[h]
  \centering
  \includegraphics[width=8cm ,height=5cm]{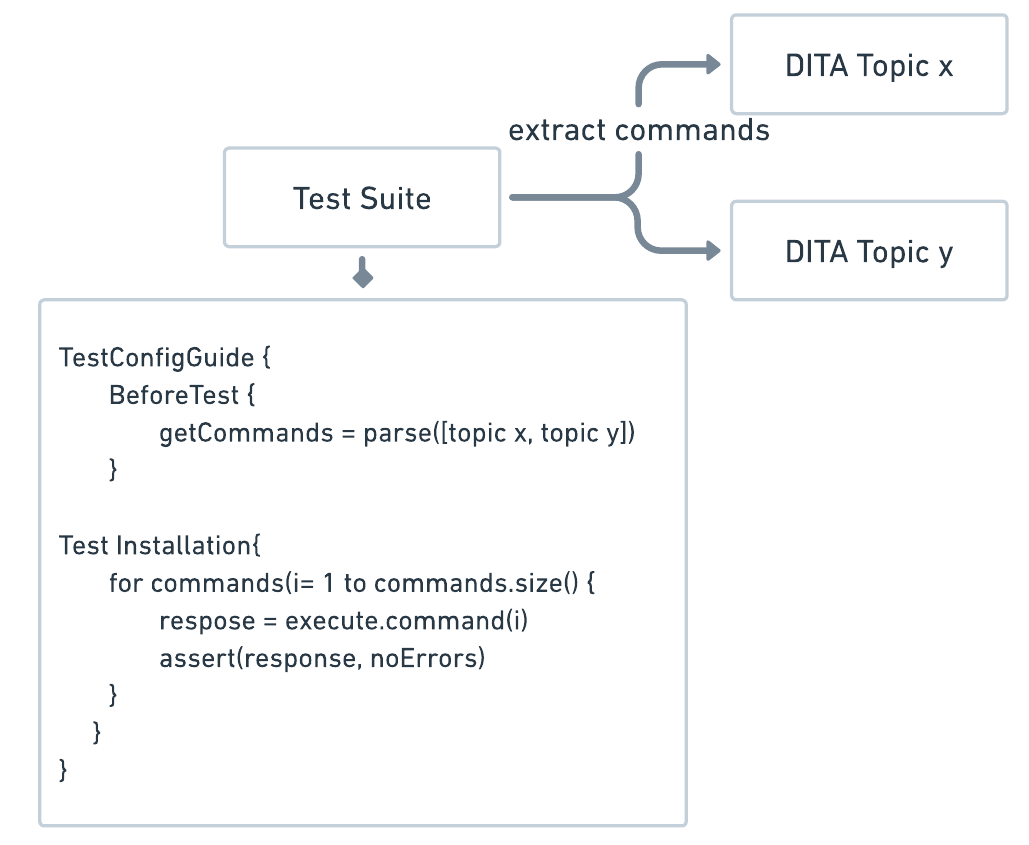}
  \caption{Extraction of commands by a test case from a DITA topic file.}
  \label{Fig:ditatopicsol}
\end{figure}

Figure~\ref{Fig:ditatopicsol} illustrates a high level implementation of a test case where the test case extracts the relevant commands from a DITA topic or topics of the ``Configuration Guide'', instead of the current practice of hard-coding commands within the test cases. Should there be a discrepancy between the documented commands and the platform/software system where the commands are being executed, the deviation would be visible due to test case failures.

\section{Discussion} \label{discussion}
\par
%The motivation for documentation debt analysis originates from  our findings in previous work on bug analysis with MultiDimEr~\cite{silva2022multidimer}, which revealed that documentation defects represented, relatively, the largest amount of bug reports of the analyzed system. 
\par
This study broadens our understanding if certain types of documentation contribute more to documentation defects (\ref{itm:RQdoctypes}). We grouped the 67 documents delivered with System A and against which defects were reported into six main documentation categories~\cite{aghajani2020software}. Deployment Guides (129 defect reports), Installation Guides (53), API References (51) and User Manuals (50) contained most of the documentation defects (see Table~\ref{table:docdefectsdistro}). A key characteristic of these documents is that they consist of steps by step instructions and commands that need to be followed and executed to achieve a certain task. We speculated therefore that the root cause for the reported defects is related to the lack of verification of the consistent co-evolution~\cite{correia2009patterns} of source code (or product features more generally) and its documentation. The solutions we describe and adapt to the context of System A are targeting the problem of having different and unreliable information sources when generating product documentation.
\par
Regarding the type of documentation defects~\ref{itm:RQdocerrors}, we found that Erroneous code examples (23), Missing documentation (36), and Outdated/Missing information (21) were the main documentation defect types, supporting our initial conjecture about their root cause, i.e. the lack of verified co-evolution. Aghajani et al.~\cite{aghajani2020software} conducted a survey (most respondents were employed at ABB, an automation technology company), with the goal of determining the relevance of the different types of documentation errors in practice. Regarding the relative importance of information content, the survey found that 30\% of the defects types related to the Information Content (What) are considered important, as opposed to only 17\% from the Information Content (How) category. Regarding the relative frequency of different defect types, the results in Table~\ref{tab:classficationResults} support the observations in the survey. The survey also revealed that Erroneous code examples was indicated as one of the top issue type based on the encountered frequency and considered important by many practitioners (59\%). Similarly, the defect types Missing user documentation, Missing documentation for a new feature/component and Outdated examples were encountered frequently, while indicating that they are perceived as important by practitioners. Our study provides therefore further evidence for the relevance of these documentation defect types and motivates the development of preventive solutions.
\par
Next, we want to discuss the suitability of the defect taxonomy for documentation debt identification by answering~\ref{itm:RQtaxSupport}. The documentation defect taxonomy~\cite{aghajani2019software} is mostly influenced by the documentation used in the development phase. For example, the defect types Erroneous code examples, Wrong code comments, Wrong translation, Missing alternative solutions can be found in source code or in developer documents rather than in customer-facing documents. In our classification we found many bug reports due to incorrect or outdated commands but we initially found it difficult to map them to any existing defect type in the taxonomy. We decided to classify such bug reports as Erroneous code examples to overcome this issue,  as our intention in this study was not to extend the taxonomy. We also found that defining document specific sub categories such as Inappropriate installation instructions, Documentation for users and Developer guidelines makes the taxonomy too overwhelming when classifying a documentation defect. A suggestion for future improvements is to define the defect types independently from the documentation types. Once the defect categories are in place, a taxonomy for different types of documentation can be defined, which makes the classification more efficient, accurate and generally applicable. 
\par
When answering~\ref{itm:RQdocCost}, we reported that the mean time for resolving a documentation defect is around 7 (external) to 10 (internal) days. Even though it is difficult to assign an absolute cost to these defects, the resolution times are high, contributing to the overall maintenance costs as well as taking up resources that could have been assigned to work on new product releases. The mean time for resolving a defect that could have been detected by our proposed automated solutions is around 11 days and could cover 87\% of the identified documentation defects from our study. We believe this is a significant improvement and paves the way for cost savings in the long run.
\par
We observed that a significant number of documentation bug reports being discovered both internally (192) within the development organization and externally (126) by the users of the system. The 40\% of externally discovered bugs is significant considering time being spent due to documentation defects in troubleshooting, the management overhead involved in the bug management process. The document types Deployment Guides, Installation Guides and User Manuals used to be manually verified by a system testing team at the department we worked with at Ericsson. However, in recent years there has been a significant investment on system test automation to reduce delivery times, which resulted in almost no verification of the documentation other than by the teams providing the documentation input. The issue here is not the actual system test automation, but the lack of linkage to the relevant documentation. 
\par
When deriving solutions as part of answering~\ref{itm:RQsolution}, we introduced three design criteria; a single and robust information source for the documentation, automation and developer friendliness. Based on these criteria we proposed two solutions, Dynamic Document Generation (DDG) and Automated Documentation Testing (ADT). The selection of a solution depends on different circumstances in different organizations. For example if it is a new project that is starting up, it would make sense to introduce DDG from the early phases of development and testing. The selection can also vary based on the type of documentation. For example, step by step instructions and the commands needed for the Installation Guides and Deployment guides are already present in CI/CD frameworks. In such circumstances, it would be more appropriate to use DDG for documentation. On the other hand, the User Guides usually contain more natural language text describing different business logic, steps to execute such business logic and boundary values for different parameters. For such documents, it may be more efficient to link the documentation to the test cases to extract boundary values required in business logic testing. 

\section{Conclusion and Future work}\label{conclusion}
\par
In this study, we emphasized that the often neglected documentation defects can be a significant contributor to the overall maintenance cost for a software development organization. We analyzed  the defects that are purely associated with documentation in a large product developed at Ericsson. We identified documentation debt by classifying the identified defects according to a taxonomy introduced from prior research~\cite{aghajani2019software}. The classification enabled us to characterize and then quantify documentation defects as a means for prioritizing solutions targeted at minimizing the occurrence of the most common documentation defects types: Erroneous code examples, Missing documentation and Outdated/Missing information. We identified three key requirements for a documentation verification system. Based on these defect types and the identified requirements, we proposed to adapt two solutions: (i) Dynamic Document Generation (DDG) and (ii) Automated Documentation Testing (ADT). The use of a  single, robust information source is the key feature of both solutions. 

\par
We presented key ideas behind the solutions such that the solutions can be implemented in different contexts. For DDG, we proposed an implementation based on Ansible, which is used extensively in the industry for installing and deploying software system in cloud native environments.

\par
In future work, we plan to implement the proposed two solutions and evaluate them in an industrial context to explore the effectiveness of the solutions and identify challenges when implementing DDG and ADT in practice. 

\subsection{Data Availability}
The 101 documentation defects and their classification is available on~\url{https://zenodo.org/record/7562614}.

\bibliographystyle{IEEEtran}
\bibliography{IEEEabrv,ref}
\end{document}